# Ordre public exceptions for algorithmic surveillance patents

Dr. iur. Alina Wernick, LL.M

The Legal Tech Lab, The Faculty of Law

The University of Helsinki, Helsinki, Finland

## Abstract

As a doctoral student, my academic relationship with Reto Hilty was characterized by the shared interest in open models of governing intellectual property law and compulsory licensing with the notion of balance as the focal point. Later, his work investigated the impact of artificial intelligence (AI) on intellectual property law[1], wherein my post-doctoral research focused on socio-legal dimensions of AI. This chapter explores the role of patent protection in algorithmic surveillance. It explores whether *ordre public* exceptions from patentability should apply to such patents, due to their potential to enable human rights violations. It concludes that in most cases, it is undesirable to exclude such patents from patentability, as the patent system is ill-equipped to evaluate the impacts of the exploitation of such technologies. Furthermore, the disclosure of such patents has positive externalities from the societal perspective, by opening the black box of surveillance for public scrutiny.

## Towards the ubiquity of algorithmic surveillance

Edward Snowden's disclosure of National Security Agent's surveillance practices in 2013 alerted policymakers, lawmakers and academics to the threats posed by state mass surveillance facilitated by the use of big data and algorithms.[2] The reach of such technology is not limited to the persons of interest of national intelligence agencies. The advances in the adoption and capabilities of artificial intelligence, facilitated wide-scale algorithmic surveillance for law enforcement or public security. Relying on machine and deep learning models, surveillance no longer needs to target specific individuals, but can sweep through and filter the studied population into groups, on the basis of their common characteristics.[3] Algorithmic surveillance is not exclusive the public sector.[4] On the one hand, the rise of social media platforms and business models based on targeted advertising has spawned the era of surveillance capitalism.[5] On the other hand, the distinction between the private and public sector has been blurred through the prominent role of the military sector in the advancement of artificial intelligence. On many instances, its applications have later been adopted for civilian use,[6] including surveillance technologies. Algorithmic surveillance may encompass cybersurveillance that intercepts information and telecommunication networks, but also cover biometric and emotion recognition systems, which may also operate remotely, without a digital interface with the targeted individual or group.[7] Algorithmic surveillance technologies can be characterized to pose risks for human rights, to have a dual use nature and to be subject to a multi-layered black-box problem.

---

[1] Lee JA, Hilty R, Liu KC (eds) (2021) Artificial Intelligence and Intellectual Property. Oxford University Press; Hilty R, Hoffman J, Scheuerer S (2021) Intellectual Property Justification for Artificial Intelligence. Draft chapter in: Lee JA, Liu KC, Hilty RM (eds) Artificial intelligence and Intellectual Property. Oxford University Press; Drexl J, Hilty R, Beneke F, Desaunettes-Barbero L, Flinck M, Globocnik J, Gonzalez Otero B, Hoffman J, Hollander L, Kim D, Richter H, Scheuerer S, Slowinski PR, Thonemann J (2019) Technical Aspects of Artificial Intelligence: An Understanding from an Intellectual Property Law Perspective. Max Planck Institute for Innovation & Competition Research Paper 19-13. http://dx.doi.org/10.2139/ssrn.3465577; Drexl J, Hilty R, Desaunettes-Barbero L, Globocnik J, Gonzalez Otero B, Hoffman J, Kim D, Kulhari S, Richter H, Scheuerer S, Slowinski PR, Widemann K (2021) Artificial Intelligence and Intellectual Property Law – Position Statement of the Max Planck Institute for Innovation and Competition of 9 April 2021 on the Current Debate. Max Planck Institute for Innovation & Competition Research Paper 21-10. http://dx.doi.org/10.2139/ssrn.3822924

[2] Lyon D (2014) Surveillance, Snowden, and big data: capacities, consequences, critique. Big Data & Society 1(2). https://doi.org/10.1177/2053951714541861

Algorithmic surveillance can be viewed to pose wide-ranging risks on human rights, starting from its capacity to violate with rights to privacy (Art. 7 CFREU) and data protection (Art. 8 CRFEU) where the limitation is not justified (Art. 52 CFREU). Deliberate oppressive use of such technologies exposes their targets to a wide range of human rights violations, starting from the interference with dignity (Art. 1 CFREU) and the mental integrity of a person (Art. 3 CFREU) and the freedoms of expression and information (Art. 11 CFREU), thought, conscience and religion (Art. 10 CFREU), assembly (Art. 12 CFREU) and movement (Art. 45 CFREU). Algorithmic surveillance may facilitate oppressive and autocratic practices.[8] Where the rule of law is weak, the targets of surveillance may be exposed to violations of due process (Art. 45-50 CFREU) be subject to risk of violations right to physical integrity of a person (Art. 3 CFREU), right to life (Art. 2 CFREU) and prohibition of torture and in human or degrading treatment or punishment (Art. 4 CFREU) as well as protection against forced labour (Art. 5 CFREU).

Algorithmic surveillance methods in themselves can also be deemed to interfere with human dignity (Art. 1 CFREU) as well as rights to physical or mental integrity of a person (Art. 2 (1) CFREU). Algorithmic surveillance systems can be deliberately designed to discriminate and target individuals belonging to certain groups (Arts. 20-24 CFREU), or they may yield discriminatory effects due to biases stemming from the data or technology.[9] For example, facial recognition technology is known to be subject to biases,[10] and to be less accurate in the recognition of ethnic minorities.[11] Furthermore, emotion recognition technology is known to have dubious epistemic foundations[12]. False positives may force an innocent person to navigate criminal proceedings.[13] At the same time, algorithmic policing often reproduces racial bias, reinforcing the histories of over policing of marginalized communities.[14]

---

[3] Kosta E (2022) Algorithmic state surveillance: Challenging the notion of agency in human rights. Regulation & Governance 16(1):212-224, pp. 214-215
[4] See Richards NM (2013) The dangers of surveillance. Harvard Law review 126(7):1934-1965
[5] Zuboff S (2019) The Age of Surveillance Capitalism: The Fight for a Human Future at the New Frontier of Power. Profile Books
[6] Crawford K (2021) The Atlas of AI: Power, Politics and the Planetary Costs of Artificial Intelligence. Yale University Press, pp. 192-193
[7] See Introna L, Wood D (2004) Picturing algorithmic surveillance: The politics of facial recognition systems. Surveillance & Society 2(2/3):177-198, p. 178
[8] See Akbari A (2022) Authoritarian Smart City: A Research Agenda. Surveillance & Society 20(4):441-449; Feldstein S (2021) The Rise of Digital Repression: How Technology is Reshaping Power, Politics, and Resistance. Oxford University; Daly A (2019) Algorithmic Oppression with Chinese Characteristics: AI Against Xinjiang's Uyghurs. In Finlay A (ed) Artificial Intelligence: Human Rights, Social Justice and Development. APC
[9] Najibi A (2020) Racial Discrimination in Face Recognition Technology. Science in the News, Harvard University. https://sitn.hms.harvard.edu/flash/2020/racial-discrimination-in-face-recognition-technology/; Castelvecchi D (2020) Is facial recognition too biased to be let loose? Nature (in online press). https://www.nature.com/articles/d41586-020-03186-4; Pfeiffer J, Gutschow J, Haas C, Möslein F, Maspfuhl O, Borgers F, Alpsancar S (2023) Algorithmic Fairness in AI: An Interdisciplinary View. Business & Information Systems Engineering 65:209-222;

Algorithmic surveillance technologies are susceptible to dual-use as "items, including software and technology, which can be used for both civil and military purposes" (Dual-Use Regulation, Art. 1(1).[15] The Dual-use Regulation was amended in 2021, introducing new rules to account for human rights threats posed by cybersurveillance technologies. These rules were initially proposed in reaction to cybersurveillance-enabled repression during the Arab Spring 2010-2012.[16] Under the Dual-use Regulation, the application of export controls for cybersurveillance is conditioned on prospect of human rights violations in the country of export.[17] The regulation defined cybersurveillance technology as items specially designed to enable the covert surveillance of natural persons by monitoring, extracting, collecting or analysing data from information and telecommunication systems."[18]

The Dual-Use Regulation's definition of cybersurveillance emphasizes its covert nature[19] Algorithmic surveillance technologies are often deliberately designed to be difficult to detect. Yet, already awareness of potentially being subject to surveillance can produce chilling effects on the enjoyment of fundamental rights.[20] Artificial Intelligence, which can be a component of cybersurveillance technology, is characterized by the black-box problem,[21] wherein the grounds for the model to produce a certain output, such the grounds for flagging a certain individual or group among those targeted with surveillance, can be difficult to explain.[22] The issues with transparency and algorithmic accountability

---

[10] Introna L, Wood D (2004) pp. 190-192

[11] Buolamwini J, Gebru T (2018) Gender Shades: Intersectional Accuracy Disparities in Commercial Gender Classification. Proceedings of Machine Learning Research 81:1-15; Noble SU (2018) Algorithms of Oppression: How Search Engines Reinforce Racism. New York University Press; Raji ID, Fried G (2021) About Face: A Survey of Facial Recognition Evaluation. arXiv:2102.00813. https://doi.org/10.48550/arXiv.2102.00813

[12] Feldman Barrett L, Adolphs R, Marsella S, Martinez AM, Pollak SD (2019) Emotional Expressions reconsidered: Challenges to Inferring Emotion from Human Facial Movements. Psychological Science in the Public Interest 20(1):1-68, p. 48.

[13] Perkowitz S (2021) The Bias in the Machine: Facial Recognition Technology and Racial Disparities. MIT Case Studies in Social and Ethical Responsibilities of Computing. https://doi.org/10.21428/2c646de5.62272586

[14] Vagle JL (2016) Tightening the ooda loop: police militarization, race, and algorithmic surveillance. Michigan Journal of Race & Law 22(1):101-138

[15] Regulation (EU) 2021/821 of the European Parliament and of the Council of 20 May 2021 setting up a Union regime for the control of exports, brokering, technical assistance, transit and transfer of dual-use items (recast), (Later, Dual-use Regulation) Art. 1 (1).

[16] Kanetake M (2019) The EU's dual-use export control and human rights risks: the case of cyber surveillance technology. Europe and the World: A Law Review 3(1):1-16, pp. 2-4, 15-16; van Daalen OL, van Hoboken JVJ, Rucz M (2023) Export control of cybersurveillance items in the new dual-use regulation: The challenges of applying human rights logic to export control. Computer law & Security Review 48

[17] Dual-use Regulation, Art. 5 (1)-(3). The use triggering human rights concerns covers the use in "internal repression and/or the commission of serious violations of human rights and international humanitarian law.". See also Dual-use Regulation, Rec. 8.

[18] Dual-use Regulation, Art. 1 (20).

[19] Dual-use Regulation

[20] Richards NM (2013) The dangers of surveillance. Harvard Law Review 126(7):1934-1965, p. 1935; Penney JW (2016) Chilling effects: Online surveillance and Wikipedia use. Berkeley Technology Law Journal 31(1): 117-182

[21] Pasquale F (2015) The Black Box Society: The Secret Algorithms that Control Money and Information. Harvard University Press.

[22] Pasquale F (2015).

are further exacerbated by the algorithmic surveillance industry's reliance on trade secret protection.[23] Furthermore, military technology is often kept in secret, also when patented.[24] The multi-layered black-box problem of algorithmic surveillance makes it difficult to scrutinize the misuses of the technology, even where national laws set conditions for their use for example through by regulating the processing of personal data and setting conditions for national intelligence through initiatives to regulate AI. Furthermore, the algorithmic surveillance can be deliberately exercised to target certain groups of population, also based on their ethnicity[25] or be exercised by foreign governments as revealed by the recent Pegasus scandal.[26] At the same time, it appears to be a thriving business.[27]

Considering the risks algorithmic surveillance technology poses for human rights, could patent law play a role in controlling its use or reducing the incentives to develop its applications? In the recent decades debates on their applicability have centred around biotechnology,[28] but the applicability of ordre public exception on applications of artificial intelligence has received little attention.[29] In this following section, I will investigate, whether patentable algorithmic surveillance technology could in some instances be exempted from patentability on the grounds of ordre public and morality (Art- 53(a) EPC).

---

[23] Matulionyte, R (2021) Reconciling Trade Secrets and AI Explainability: Face Recognition Technologies as a Case Study (November 30, 2021). R Matulionyte, 'Reconciling Trade Secrets and Explainable AI: face recognition technology as a case study' (2022) 44(1) European Intellectual Property Review 46: https://ssrn.com/abstract=3974221

[24] Matthews D, Ostapenko H (2023) The War in Ukraine Raises Questions About Patents for Secret Inventions. GRUR International. https://doi.org/10.1093/grurint/ikad042

[25] Kelion L (2021) Huawei patent mentions use of Uighur-spotting tech. BBC (in online press). https://www.bbc.com/news/technology-55634388; Harwell D, Dou E (2020) Huawei tested AI software that could recognize Uighur minorities and alert police, report says. The Washington Post (in online press). https://www.washingtonpost.com/technology/2020/12/08/huawei-tested-ai-software-that-could-recognize-uighur-minorities-alert-police-report-says/

[26] Joyner E (2022) The EU Watergate? Pegasus spyware scandal grows. DW (in online press). https://www.dw.com/en/eu-watergate-the-pegasus-spyware-scandal-keeps-spreading/a-63687981

[27] Feldstein S, Kot B (2023) Why Does the Global Spyware Industry Continue to Thrive? Trends, Explanations, and Responses. Carnegie working paper. Available at: https://carnegieendowment.org/2023/03/14/why-does-global-spyware-industry-continue-to-thrive-trends-explanations-and-responses-pub-89229; Report Insight Consulting (2023) Facial Recognition Market Set For Significant Expansion, Exceed to US$16.5 Billion by 2030 Projects Reports Insights Study. GlobeNewswire. https://www.globenewswire.com/en/news-release/2023/03/16/2628417/0/en/Facial-Recognition-Market-Set-For-Significant-Expansion-Exceed-to-US-16-5-Billion-By-2030-Projects-Reports-Insights-Study.html: Grand View Research (2023) Emotion Detection And Recognition Market Size, Share, & Trend Analysis Report By Component, By Tools, By Technology, By Application, By End Use Vertical, By region And Segment Forecasts, 2022-2030. https://www.grandviewresearch.com/industry-analysis/emotion-detection-recognition-market-report

[28] See for example Straus, J. (2013) Ordre public and morality issues in patent eligibility. In *Intellectual property in common law and civil law*, pp. 19-49. Edward Elgar Publishing; Liddell, K (2012). Immorality and patents: the exclusion of inventions contrary to ordre public and morality. *New Frontiers in the Philosophy of Intellectual Property* 18 (2012): 140; Prifti, V (2019) "The limits of "ordre public" and "morality" for the patentability of human embryonic stem cell inventions." The Journal of World Intellectual Property 22(1-2): 2-15; Nordberg, A (2020), Patents, Morality and Biomedical Innovation in Europe: Historical Overview, Current Debates on Stem Cells, Gene Editing and AI, and de lege ferenda Reflections (June 13, 2019). Forthcoming in Daniel Gervais (ed) Fairness, Morality and Ordre Public (Edward Elgar, 2020), Available at SSRN: https://ssrn.com/abstract=3540234.; Matthews, D, Minssen T, and Nordberg A (2022) Balancing innovation,'ordre public'and morality in human germline editing: A call for more nuanced approaches in patent law. European Journal of Health Law 29(3-5): 562-588.

[29] Cf. Spranger, TM (2023) Brain Patents as a Legal or Societal Challenge?. IIC 54: 268–275, discussing the application of ordre public exception on neurotechnology patents.

## Ordre public exceptions for algorithmic surveillance technology

According to Art. 27 (2) TRIPS, member states may exclude from patentability inventions, the prevention within their territory of the commercial exploitation of which is necessary to protect *ordre public* or morality, including to protect human, animal or plant life or health or to avoid serious prejudice to the environment, provided that such exclusion is not made merely because the exploitation is prohibited by their law.

The EPC poses slightly narrower conditions for the application of the *ordre public* and morality exception.[30] Article 53 (a) EPC stipulates that European patents shall not be granted in respect of inventions the commercial exploitation of which would be contrary to "ordre public" or morality; such exploitation shall not be deemed to be so contrary merely because it is prohibited by law or regulation in some or all of the Contracting States.[31] Under the EPC, that exception is "likely to be invoked only in rare and extreme cases." [32]

The concept of ordre public does not have a generally accepted definition. [33] Generally, civil and human rights serve as the foundation of the public order in Europe.[34] Furthermore, the concept of "ordre public" is viewed to cover "the protection of public security and the physical integrity of individuals as part of society." Inventions the exploitation of which is likely to breach "public peace and social order (for example, through acts of terrorism)" would fall under the prohibition of Art. 53(a) .[35] According to the EPO's Guidelines for examination, the exception is meant "to deny protection to inventions likely to induce riot or public disorder, or to lead to criminal or other generally offensive behaviour."[36] Since algorithmic surveillance technologies are often developed for the specific purpose of enhancing security and safety purposes, they are unlikely to be considered contrary to *ordre public*, even where their exploitation could potentially lead to human rights violations, such as oppression of dissidents.

This conclusion is reinforced by the fact that the exception of Art. 53 (a) EPC does not apply to technologies with a potential for misuse. According to the Guidelines "the mere possibility of abuse of an invention is not sufficient to deny patent protection if the invention can also be exploited in a way which does not and would not infringe "ordre public" and morality". [37] Same applies also to uses that may be penalized.[38] As a consequence, Art. 53(a) EPC would not apply to the majority of algorithmic surveillance patents. No technical solution that has a potential to be used both in a lawful way and in a manner that could interfere with public security, or which could be used for discriminatory or oppressive purposes can be excluded from patentability on grounds of Art. 53(a)

Would apply also to technologies that are accurate detecting sensitive data about a person, such as their race or ethnicity? Could algorithmic surveillance patents, that for example, that target minorities, be patentable?.³⁹ Art. 27 (2) TRIPS is viewed to give considerable leeway for the member state's subjective interpretation.⁴⁰ The concept of morality is also deemed culture and time-specific.⁴¹ The Examination Guidelines qualify the criterion as whether it is probable that the public in general would regard the invention as so abhorrent that the grant of patent rights would be inconceivable".⁴² The Boards of Appeal have anchored the concept of morality to the European culture:

> The concept of morality is related to the belief that some behaviour is right and acceptable whereas other behaviour is wrong, this belief being founded on the totality of the accepted norms which are deeply rooted in a particular culture. For the purposes of the EPC, the culture in question is the culture inherent in European society and civilisation. Accordingly, under Article 53(a) EPC, inventions the exploitation of which is not in conformity with the conventionally-accepted standards of conduct pertaining to this culture are to be excluded from patentability as being contrary to morality.⁴³

---

[30] Haugen HM (2009) Human Rights and TRIPS Exclusion or Exception Provision. The Journal of World Intellectual Property 11(5-6):345-374, p. 349

[31] The Rule 28 of the Implementing Regulations specifies the meaning of the exception with respect to biotechnological processes, such as human cloning, which are prohibited from patenting.

[32] Guidelines for Examination, Part G II [4.1].

[33] Correa CM, "Patent Rights." In Correa CM and Yusuf Aa (eds) Intellectual Property and International Trade: The TRIPS Agreement, Kluwer Law International, 2016., p. 267

[34] Slaughterhause T-0149/11 [2].

[35] Plant Genetic Systems N.V., et al T 0356/93 [5].

[36] Guidelines for Examination, Part G II [4.1].

[37] Guidelines for Examination, Part G II [4.1.2].

[38] Guidelines for Examination, Part G II [4.1].

The fact that algorithmic surveillance technology can detect sensitive categories of data alone does not warrant that it is contradicting with public morality. In fact, this feature may be essential for the technology function accurately.⁴⁴ However, patents that aim at blatant interferences with public morals, such as deliberately targeting certain ethnic groups with algorithmic surveillance for the purposes of policing and, for example, integrating such functions into weapons should trigger the assessment on excludability of such patent on moral grounds.⁴⁵ Normatively, such exclusions would be rooted in the protection of human life and dignity as well as European cultural and historical background, including the genocide of several ethnic populations and minorities during the *Holocaust.*

It is more difficult to draw the line where the ontologies behind surveillant applications are questionable and can lead to severe inaccuracies or discriminatory effects. The evaluation of these shortcomings would be more appropriate to carry out in connection with the evaluation of the fulfilment of the inventive step (Art. 56 EPC).

It is essential, that the contradiction with public order or morality is evaluated with a view of commercial exploitation of the patent. However, this creates a temporal tension with the other criteria of patentability. Whereas the conditions for novelty and inventive step focus on the state of the art at the date of filing or priority, the ordre public and morality exception requires the patent examiner to foresee the how the patent in question is commercialized.[46] It may be infeasible to predict the impact of technology at its very early stage of development,[47] as undesirable effects of AI can be very context-specific[48] and may emerge throughout the life-cycle of the technology.[49]

---

³⁹ This could occur by directly processing sensitive categories of data (Art. 9 GDPR). Alternatively, the solution could rely on processing personal or non-personal data to draw "'high-risk inferences,' meaning inferences drawn through Big Data analytics that are privacy-invasive or reputation-damaging, or have low verifiability in the sense of being predictive or opinion-based while being used for important decisionsWachter, S & Mittelstadt, B (2019) Right to reasonable inferences: re-thinking
data protection law in the age of big data and AI. Columbia Business Law Review,
2019(2):494-620, p. 500.
⁴⁰ Malbon J, Lawson C, Davison M (2014) The WTO Agreement on Trade-Related Aspects of Intellectual Property Rights: A Commentary. Edward Elgar Publishing, p. 434
⁴¹ Correa, p. 267.
⁴² Guidelines for Examination, Part G II [4.1].
⁴³ Plant Genetic Systems N.V., et al T 0356/93 [6].
⁴⁴ Plenke, M (2015) The Reason This "Racist Soap Dispenser" Doesn't Work on Black Skin. Mic. https://www.mic.com/articles/124899/the-reason-this-racist-soap-dispenser-doesn-t-work-on-black-skin
⁴⁵ For example, several Chinese patents related to fascial recognition were found to explicitly enable the tracking of the Uyghur population. See Healy, C (2021) Uyghur Surveillance & Ethnicity Detection Analytics in China. Expert Report Presented to the Uyghur Tribunal IPVM, describing a number of patents explicitly distinguishing Uyghurs from other ethnicities. After being subject to public attention, one of the patents discussed was subsequently amended and two were withdrawn. IPVM (2021) Patenting Uyghur Tracking – Huawei, Megvii and More. https://ipvm.com/reports/patents-uyghur

The mere fact that the exploitation of the patented technology is legally prohibited does not amount application of the *ordre public* exception, as it may be manufactured for export.[50] The exception form patentability is determined independently from other legislation that could prohibit or limit the use algorithmic surveillance technologies a member state,[51] such as the prohibited applications of artificial intelligence under the Artificial Act Proposal.[52] However, wherein the use of technology is associated with a criminal offence, could contribute to the analysis on whether the exploitation of the technology could violate the public order. On the same basis, it could limit patentability of certain algorithmic surveillance patents, even where their use is expressly permitted in the relevant country.

## Expansion of the ordre public exception as a way forward?

The narrow scope of the *ordre public* and morality exception in the face of emerging technologies has been criticized for lack of accounting for future applications of the technology, coherence, social and political context as well as the excessive burden of evidence posed on the opponent.[53] To address the shortcomings of the ordre public and morality exception, Grosse Ruse-Khan has proposed a stricter test that prima facie, the predominant application of the invention is "helpful and not harmful".[54] Pila supports a more extensive normative and procedural adjustment of the exception by grounding application to the precautionary principle. In practice, this would mean carrying out a risk assessment process, which involves a wider set of stakeholders that patent opposition procedure entails, such as government, industry, and consider a wider set of "social, cultural ethical, political significance if patenting an emergent technology". The risk assessment would be carried out by the patent office with respect to patents on emerging technologies.[55]

While I agree on relevance role of the precautionary principle as well as risk and impact assessments in the regulation and governance of emergent technologies, I do not think that the patent office should be vested with such a responsibility. In the case of algorithmic surveillance technology, the assessment of individual patent applications may give limited account on the impact as well as risks associated with the technology, especially if it contains elements and components that are not patentable, either on the grounds of being computer programs (Art 52(2)(c)) or mathematical methods (Art. 52(2)(a) or where the impact or risk stems from the use of numerous patentable components. Algorithmic surveillance technologies build upon a constellation of patents that support the deployment of wider surveillance platforms, such as the solution provided by Palantir.[56] Furthermore, algorithmic surveillance solutions rely on general purpose technologies[57], such as sensors and cameras, which would not trigger the *ordre public* exception.

---

[46] See also Pila J (2020) Adapting the ordre public and morality exclusion of European patent law to accommodate emerging technologies. Nature biotechnology 38(5):555-557, p. 555.
[47] Grosse Ruse-Khan, H (2023) Does IP improve the world?. In Improving Intellectual Property (pp. 492-502). Edward Elgar Publishing, p. 503.
[48] Birhane A (2021) Algorithmic injustice: a relational ethics approach. Patterns 2(2).
[49] Suresh H, Guttag J (2021) A framework for understanding sources of harm throughout the machine learning life cycle. Paper presented at the EAAMO '21 conference on Equity and Access in Algorithms, Mechanisms, and Optimization, 5-9 October, 2021. Association for Computing Machinery, New York, US.
[50] Guidelines for Examination, Part G II [4.1.1].
[51] However, wherein the use of technology is associated with a criminal offence, could be viewed to contribute to the analysis on whether the exploitation of the technology could violate the public order. See Guidelines for Examination, Part G II [4.1].
[52] Proposal for a regulation of the European Parliament and of the Council laying down harmonised rules on artificial intelligence (Artificial intelligence act) and amending certain union legislative acts, COM/2021/206 final, Title II Prohibited artificial intelligence practices.
[53] Pila p. 555.
[54] Grosse Ruse-Khan, p. 503.
[55] Pila, p. 556.

It also would not be desirable to condition the exception the invention carrying some degree of risk to interfere with fundamental rights. The fact that certain technology may interfere with the enjoyment of fundamental rights, does not mean that interference would not be deemed proportionate considering the justification for interference in the given application context of the technology.[58]

One may also ask whether the applicability of the *ordre public* exception would diminish some of the advantages that the patent system provides with respect to controlling for negative effects of the patented invention. First, patents could be applied to with the aim of preventing others from using the technology that can have harmful effects. For example, the inventor of the atomic bomb, Leo Szilard, had applied patent a patent for his invention with the intention of preventing others from using the dangerous invention.[59] The owner of a patent on a dual use technology may also want to benefit from the possibility to select their licensees to avoid its undesirable applications.

The exercise of the precautionary principle and obligation to carry out or oversee risk and impact assessment should be vested with another agency also for another reason. Furthermore, limiting the patentability of algorithmic surveillance technologies could drive their developments towards further reliance on trade secrecy and further limits the possibilities to oversee excesses, misuses and unwanted effects associated with their deployment. Instead, it may be fruitful to explore the role that patent law can play in making algorithmic surveillance technology and its use more transparent by virtue of mandatory patent disclosure.

In sum, *ordre public* and morality exception is not effective in controlling for undesirable effects of algorithmic surveillance. Aside from blatantly immoral applications there is no reason to expand the applicability of the exception vis-à-vis algorithmic surveillance technology or oblige patent offices to carry out risk assessments as proposed by Pila. In fact, doing so could have detrimental effects on containing undesirable effects of algorithmic cybersurveillance. Instead, the patent system could contribute to the multi-layered black-box problem associated with algorithmic surveillance technologies through its disclosure function.

### Patents disclosure as means of enhancing algorithmic transparency.

With the exception of military patents, patent protection is conditioned on the disclosure of the patent application (Art. 29 (1) TRIPS; Art. 93 (1) EPC). Emergent research suggests that patent data can be used to track and understand the developments in cybersurveillance technology. For example, Iliadis and Acker employed the method of computational topic modelling on Palantir's patents to understand how its surveillance platform operates. Their analysis revealed the company's strategic objective of "all-encompassing data integration work" across clients, data subjects technological interfaces and diverse datasets.[60] Wright et al used the method of topic modelling on Chinese patents associated smart cities, to uncover key themes among the patents, many of which were associated with policing.[61] Algorithmic analysis of patent data has informed also legal research on human rights risks associated with neurotechnology.[62] Qualitative research on patents allows for the identification of imaginaries associated with algorithmic surveillance technologies.[63]

---

[56] Iliadis A & Acker A (2022). The seer and the seen: Surveying Palantir's surveillance platform. The Information Society, 38(5):334-363.
[57] Van Daalen et al p. 10.
[58] Art. 52 CFREU.
[59] Nuclear Weapon Archive (1997) Invention and Discovery: Atomic Bombs and Fission, Leo Szilard and the Invention of the Atomic Bomb. https://nuclearweaponarchive.org/Usa/Med/Discfiss.html

These studies show that beyond the classical function of patent disclosure, it also creates a positive externality of informing the public on the imaginaries, trajectories and power dynamics attached to the technology development. Furthermore, statistical and algorithmic methods can assist in gaining insights from patent data for the purposes of identifying key developments and actors in the field.

It must be, of course, acknowledged that fact that an invention is patented, does not mean that it will ever be implemented or commercialized. The fact that a patent raises concerns of negative effects of cybersurveillance, does not mean that the patent is actually exercised in a questionable manner. Yet, patents present rich, public data source [64] for analysing emergent developments withing specific technological sectors. It is of particular value to inform policymaking and regulation in the sectors subject to multi-layered opacity, as in the case of algorithmic surveillance.

Algorithmic surveillance technologies are likely to result from assemblages of patented technologies and software.[65] The assemblages, surveillant features and unjustifiable applications may materialize long time after the patenting of relevant technology. For this reason, patent offices are not the right agencies to be vested with responsibility to predict the future risks associated with technology. However, they could be equipped with an obligation to signal to responsible authority where patented technologies and patenting raise concerns of possibly problematic, future uses. For example, to foster interdisciplinary learning on the human rights risks and implications of neurotechnology, Spranger proposes to patent offices to notify human rights agencies on patents that triggered the to an ordre public examination, but were not denied patent protection on the grounds of having also non-harmful uses.[66] In the context of algorithmic surveillance patents, the relevant authorities could be those vested with powers oversee the enforcement of as well as to propose revisions to Artificial Intelligence Act. A possibility to grant such authorities access to yet-to-be published patents applications, could also allow them to develop timely responses emerging applications algorithmic surveillance technologies, for through risk and impact assessments or instruments such as regulatory sandboxes.

## Conclusion

The development, use and functioning of algorithmic surveillance technologies are subject to multilayered black-box problem.[67] Being dual-use technology, they can be exercised in a manner that can pose risks to human rights. Reliance on artificial intelligence to carry out surveillance, for example to identify, track and identify people or groups raises further human rights risks. This chapter explored whether such technologies could on some occasions be rendered unpatentable on the grounds of the ordre public and morality exception. Although human rights inform the normative basis of the evaluation of the applicability of the exception, it does not apply to technology that has potential to be misused. Algorithmic surveillance technology applications, such as facial recognition technologies that single out individuals or groups on the basis of their ethnicity or similar high-risk inferences[68], could trigger the review of the excludability of the patent application on the grounds of morality. The line could be difficult to draw, as identifying those characteristics may also be necessary for preventing false positives upon applying the technology.

---

[60] Iliadis A & Acker A (2022) The seer and the seen: Surveying Palantir's surveillance platform. The Information Society, 38(5):334-363, p. 357.
[61] Wright J, Weber V & Walton G (submitted manuscript, 2023): Identifying emerging human rights implications in Chinese smart cities via machine-learning aided patent analysis.
[62] Spranger, 2023.
[63] Shapiro, A. (2020). 'Embodiments of the invention': Patents and urban diagrammatics in the smart city. Convergence, 26(4), 751-774.
[64] See Iliadis & Acker.
[65] See Haggerty, KD & Ericson R V (2000). The surveillant assemblage. The British journal of sociology, 51(4), 605-622, p. 608.
[66] Spranger, 274.
[67] See Pasquale.

While ordre public exception fails to account for the application context and future uses of the technology[69] patent offices are unlikely to be the right forum to account for human rights risks of algorithmic surveillance technologies. Algorithmic surveillance builds upon assemblages of patentable and unpatentable technologies, part of which are protected with trade secrets. The impact of such technologies cannot be assessed by looking on an individual patent in isolation. Furthermore, such technologies may be deliberately patented to enable their responsible use or prevent the exploitation of such technology. Finally, expanding the ordre public exception is likely to only drive the developers of such technologies to rely on trade secrecy and exacerbating the black-box problem surrounding their use. Patent data is already being used to make sense of developments in the algorithmic surveillance technologies therefore patent offices should focus on the value that disclosure function bring to research, regulation and policy-making on algorithmic surveillance technologies. Possible amendments to the patent grant procedure should further enhance the value of the disclosure to the society.

---

[68] See Wachter & Mittelstand.
[69] See Pila; Grosse Ruse-Khan.